# Mapping the Ethics of Generative AI
## A Comprehensive Scoping Review

Thilo Hagendorff
thilo.hagendorff@iris.uni-stuttgart.de
University of Stuttgart
Interchange Forum for Reflecting on Intelligent Systems

**Abstract** – The advent of generative artificial intelligence and the widespread adoption of it in society engendered intensive debates about its ethical implications and risks. These risks often differ from those associated with traditional discriminative machine learning. To synthesize the recent discourse and map its normative concepts, we conducted a scoping review on the ethics of generative artificial intelligence, including especially large language models and text-to-image models. Our analysis provides a taxonomy of 378 normative issues in 19 topic areas (accessible [online](#)) and ranks them according to their prevalence in the literature. The study offers a comprehensive overview for scholars, practitioners, or policymakers, condensing the ethical debates surrounding fairness, safety, harmful content, hallucinations, privacy, interaction risks, security, alignment, societal impacts, and others. We discuss the results, evaluate imbalances in the literature, and explore unsubstantiated risk scenarios.

## 1 Introduction

With the rapid progress of artificial intelligence (AI) technologies, the ethical reflection thereof is constantly facing new challenges. From the advent of deep learning for powerful computer vision applications[1], to the achievement of superhuman-level performance in complex games with reinforcement learning (RL) algorithms[2], and large language models (LLMs) possessing complex reasoning abilities[3,4], new ethical implications have arisen at extremely short intervals in the last decade. Likewise, while the discipline of AI ethics began by being mostly a reactive endeavor erecting normative principles for entrenched AI technologies[5,6], it became increasingly proactive with the prospect of harms through misaligned artificial general intelligence (AGI) systems. During its evolution, AI ethics underwent a practical turn to explicate how to put principles into practice[7,8]; it diversified into alternatives for the principle-based approach, for instance by building AI-specific virtue ethics[9,10]; it received criticism for being inefficient, useless, or whitewashing[11–14]; it became increasingly transferred into proposed legal norms like the AI Act of the European Union[15,16]; and it became accompanied by two new fields dealing with technical and theoretical issues alike, namely AI alignment and AI safety[17,18]. Both domains have a normative grounding and are devoted to preventing harm or even existential risks stemming from generative AI systems.

On the technical side of things, variational autoencoders[19], flow-based generative models[20,21], or generative adversarial networks[22] were early successful generative models, supplementing discriminatory machine learning architectures. Later, the transformer architecture[23] as well as diffusion models[24] boosted the performance of text and image generation models and made them adaptable to a wide range of downstream tasks. This development is covered by the idea of foundation models[25]. However, due to the lack of user-friendly graphical user interfaces, dialog optimization, and output quality, generative models were underrecognized in the wider public. This changed with the advent of models like ChatGPT, Gemini, Stable Diffusion, or Midjourney, which are accessible through natural language prompts and easy-to-use browser interfaces[26–28]. The next phase will see a rise in multi-modal models, which are similarly user-friendly and combine the processing and generation of text, images, and audio along with other modalities, such as tool use[29,30]. In sum, we define the term "generative AI" as comprising large, foundation, or frontier models, capable of transforming text to text, text to image, image to text, text to code, text to audio, text to video, or text to 3D[31].

The swift innovation cycles in machine learning and the plethora of related normative research works in ethics, alignment, and safety research make it hard to keep track. To remedy this situation, meta-studies provided synopses of AI policy guidelines[32], sociotechnical harms of algorithmic systems[33], values in machine learning research[34], risks of specific applications like language models[35], occurrences of harmful machine behavior[36], impacts of generative AI on cybersecurity[37], the evolution of research priorities in generative AI[38], and many more. While these meta-studies render the research community a tremendous service, no such meta-study exists that targets the assemblage of ethical issues associated with the latest surge of generative AI applications at large.

As a scoping review, this study is supposed to close this gap and to provide a practical overview for scholars, AI practitioners, policymakers, journalists, as well as other relevant stakeholders. Based on a systematic literature search and coding methodology, we distill the body of knowledge on the ethics of generative AI, synthesize the details of the discourse, map normative concepts, discuss imbalances, and provide a basis for future research and technology governance. The complete taxonomy, which encompasses all ethical issues identified in the literature, is available online [here](.).

## 2 Methods

We conducted a scoping review[39] with the aim of covering a significant proportion of the existing literature on the ethics of generative AI. Throughout the different phases of the review, we followed the PRISMA (Preferred Reporting Items for Systematic Reviews and Meta-Analyses) protocol[40,41]. In the first phase, we conducted an exploratory reading of definitions related to generative AI to identify key terms and topics for structured research. This allowed us to identify 29 relevant keywords for a web search. We conducted the search using a Google Scholar API with a blank account, avoiding the influence of cookies, as well as the arXiv API. We also scraped search results from PhilPapers, a database for publications from philosophy and related disciplines like ethics. Next to that, we used the AI-based paper search engine Elicit with 5 tailored prompts. For details on the list of keywords as well as prompts, see Appendix A. We collected the first 25 (Google Scholar, arXiv, PhilPapers) or first 50 (Elicit) search results for every search pass, which resulted in 1.674 results overall, since not all search terms yielded 25 hits on arXiv or PhilPapers. In terms of the publication date, we included papers from 2021 onwards. Even though this is a brief timeframe, it encompasses the full phase of emergent excitement surrounding generative AI. This was initially sparked by the release of OpenAI's DALL-E[42] in 2021 and later intensified by the tremendous popularity of ChatGPT[26] in 2022.

We deduplicated our list of papers by removing string-wise identical duplicates as well as duplicate titles with a cosine similarity above 0.8 to cover title pairs which possess slight capitalization or punctuation differences. Eventually, we retrieved 1120 documents for title and abstract screening. Of those, 162 met the eligibility criteria for full text screening (see Appendix B). Based on them, we used citation chaining to identify additional records by sifting through the reference lists of the original papers until no additional publication could be identified (see Appendix C). Furthermore, we monitored the literature after our initial search was performed to retrieve additional relevant documents (see Appendix C). For the latter two approaches, we implemented the limitation that we only considered overview papers, scoping reviews, literature reviews, or taxonomies. Eventually, the identification of further records resulted in 17 additional papers. In sum, we identified 179 documents eligible for the detailed content analysis (see Appendix D). The whole process is illustrated in the flowchart in Figure 1.

For the paper content analysis and annotation, we used the data analysis software NVivo (version 14.23.2). In the initial coding cycle, all relevant paper texts were labelled paragraph by paragraph through a bottom-up approach deriving concepts and themes from the papers using inductive coding[43]. We only coded arguments that possess an implicit or explicit normative dimension, meaning statements about what ought to be, discussions of harms, opportunities, risks, norms, chances, values, ethical principles, or policy recommendations. We did not code purely descriptive content or technical details unrelated to ethics. Moreover, we did not code arguments if they did not pertain to generative AI but traditional machine learning methods like classification, prediction, clustering, or regression techniques. Additionally, we did not annotate paper appendices. New codes were created once a new normative argument, concept, principle, or risk was identified until theoretical saturation was reached over all analyzed papers.

Once the initial list of codes was created by sifting through all sources, the second coding cycle started. Coded segments were re-checked to ensure consistency in code application. All codes were reviewed, discrepancies resolved, similar or redundant codes clustered, and high-level categories created. Eventually, the analysis resulted in 378 distinct codes.

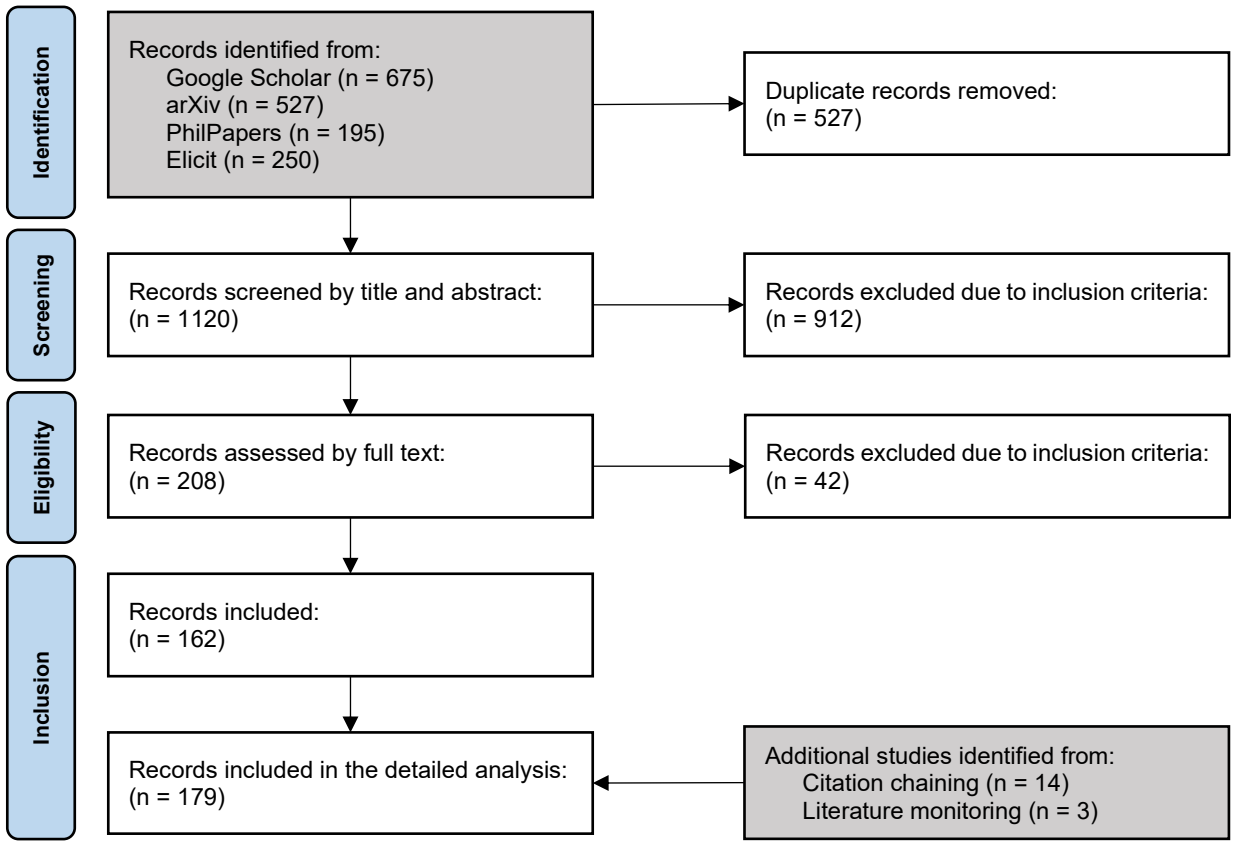

*Figure 1 - Flow diagram illustrating the paper selection process.*

## 3 Results

Previous meta-studies on AI ethics guidelines[6,32,44] congruently found a set of reoccurring paramount principles for AI development and use: transparency, fairness, security, safety, accountability, privacy, and beneficence. However, these studies were published before the excitement surrounding generative AI[26,42]. Since then, the ethics discourse underwent significant changes, reacting to the new technological developments. Our analysis of the recent literature revealed that new topics emerged, comprising issues like jailbreaking, hallucination, alignment, harmful content, copyright, models leaking private data, impacts on human creativity, and many more. Concepts like trustworthiness or accountability lost importance, while others became even more prevalent, especially fairness and safety. Still other topics remained very similar, for instance discussions surrounding sustainability or transparency. In sum, our review revealed 19 distinct clusters of ethics topics, all of which will be discussed in the following, in descending order of importance (see Figure 1). The complete taxonomy comprising all ethical issues can be accessed [here](#).

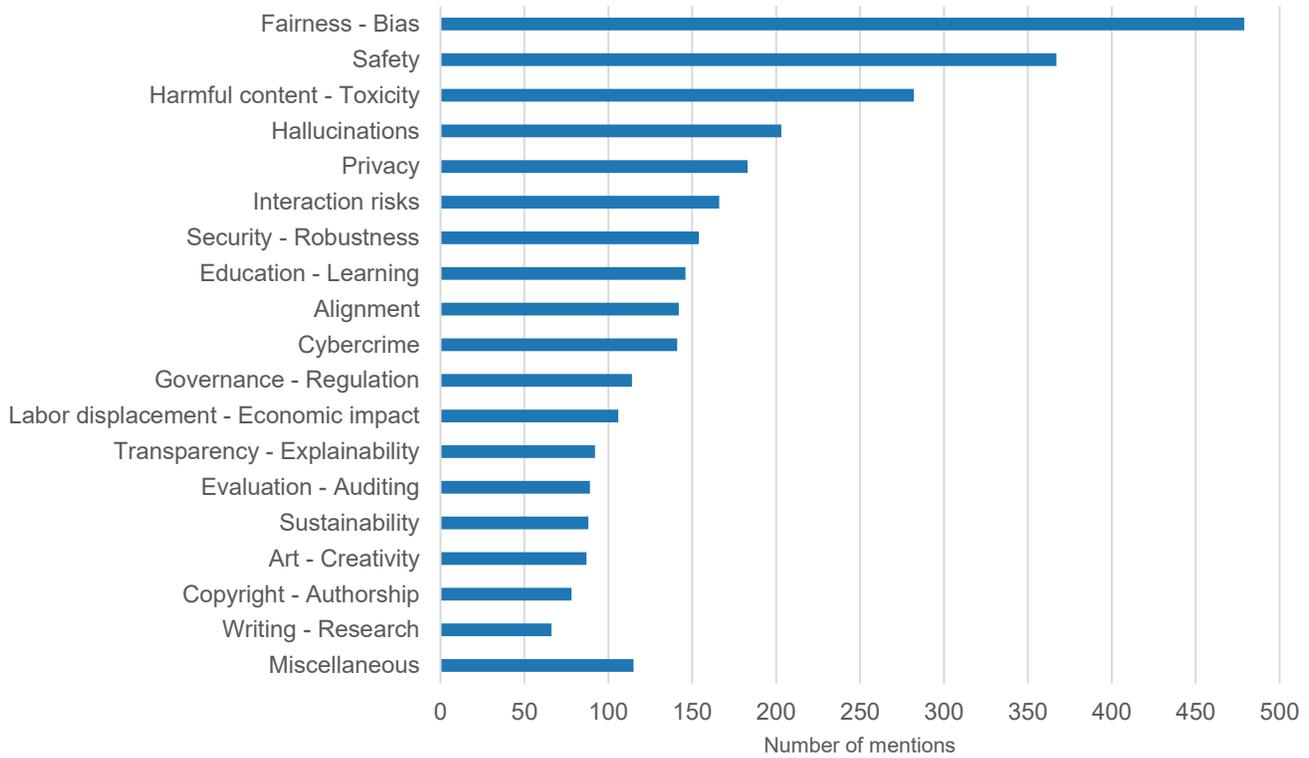

*Figure 1 - Overview of identified topic clusters and their quantitative prevalence in the literature.*

**Fairness - Bias**. Fairness is, by far, the most discussed issue in the literature, remaining a paramount concern especially in case of LLMs and text-to-image models[35,45–47]. This is sparked by training data biases propagating into model outputs[48], causing negative effects like stereotyping[33,35], racism[49], sexism[50], ideological leanings[47], or the marginalization of minorities[51]. Next to attesting generative AI a conservative inclination by perpetuating existing societal patterns[52], there is a concern about reinforcing existing biases when training new generative models with synthetic data from previous models[53]. Beyond technical fairness issues, critiques in the literature extend to the monopolization or centralization of power in large AI labs[25,54–56], driven by the substantial costs of developing foundational models. The literature also highlights the problem of unequal access to generative AI, particularly in developing countries or among financially constrained groups[35,47,57,58]. Sources also analyze challenges of the AI research community to ensure workforce diversity[59]. Moreover, there are concerns regarding the imposition of values embedded in AI systems on cultures distinct from those where the systems were developed[51,60].

**Safety**. The second prominent topic in the literature, as well as a distinct research field in its own right, is AI safety[17]. A primary concern is the emergence of human-level or superhuman generative models, commonly referred to as AGI, and their potential existential or catastrophic risks to humanity[54,61–63]. Connected to that, AI safety aims at avoiding deceptive[36,64] or power-seeking machine behavior[54,65,66], model self-replication[54,67], or shutdown evasion[54,67]. Ensuring controllability[65], human oversight[68], and the implementation of red teaming measures[54,69] are deemed to be essential in mitigating these risks, as is the need for increased AI safety research[54,67] and promoting safety cultures within AI organizations[54] instead of fueling the AI race[54]. Furthermore, papers thematize risks from unforeseen emerging capabilities in generative models[68,70], restricting access to dangerous research works[71,72], or pausing AI research for the sake of improving safety or governance measures first[61,73]. Another central issue is the fear of weaponizing AI or leveraging it for mass destruction[54], especially by using LLMs for the ideation and planning of how to attain, modify, and disseminate biological agents[74,75]. In general, the threat of AI misuse by malicious individuals or groups[47], especially in the context of open-source models[68], is highlighted in the literature as a significant factor emphasizing the critical importance of implementing robust safety measures.

**Harmful content - Toxicity**. Generating unethical, fraudulent, toxic, violent, pornographic, or other harmful content is a further predominant concern, again focusing notably on LLMs and text-to-image models[25,33,35,53,58,69,76–80]. Numerous studies highlight the risks associated with the intentional creation of disinformation[35], fake news[77], propaganda[80], or deepfakes[47], underscoring their significant threat to the integrity of public discourse and the trust in credible media[53,81]. Additionally, papers explore the potential for generative models to aid in criminal activities[82], incidents of self-harm[72], identity theft[35], or impersonation[83]. Furthermore, the literature investigates risks posed by LLMs when generating advice in high-stakes domains such as health[84], safety-related issues[85], as well as legal or financial matters[86].

**Hallucinations**. Significant concerns are raised about LLMs inadvertently generating false or misleading information[33,35,47,65,69,77,87–95], as well as erroneous code[93,96]. Papers not only critically analyze various types of reasoning errors in LLMs[87] but also examine risks associated with specific types of misinformation, such as medical hallucinations[97]. Given the propensity of LLMs to produce flawed outputs accompanied by overconfident rationales[93] and fabricated references[86], many sources stress the necessity of manually validating and fact-checking the outputs of these models[92,98,99].

**Privacy**. Generative AI systems, similar to traditional machine learning methods, are considered a threat to privacy and data protection norms[35,47,100,101]. A major concern is the intended extraction or inadvertent leakage of sensitive or private information from LLMs[72,77,100,102,103]. To mitigate this risk, strategies such as sanitizing training data to remove sensitive information[103] or employing synthetic data for training[104] are proposed. Furthermore, growing concerns over generative AI systems being used for surveillance purposes emerge[35,56]. To safeguard privacy, papers stress the importance of protecting sensitive and personal data transmitted to AI operators[84,105,106]. Moreover, they are urged to avoid privacy violations during training data collection[56,77,107].

**Interaction risks.** Many novel risks posed by generative AI stem from the ways in which humans interact with these systems[35]. For instance, sources discuss epistemic challenges in distinguishing AI-generated from human content[79]. They also address the issue of anthropomorphization[108], which can lead to an excessive trust in generative AI systems[109]. On a similar note, many papers argue that the use of conversational agents could impact mental well-being[47,109] or gradually supplant interpersonal communication[78], potentially leading to a dehumanization of interactions[47]. Additionally, a frequently discussed interaction risk in the literature is the potential of LLMs to manipulate human behavior[18,36,110] or to instigate users to engage in unethical or illegal activities[35].

**Security - Robustness**. While AI safety focusses on threats emanating from generative AI systems, security centers on threats posed to these systems[94,111]. The most extensively discussed issue in this context are jailbreaking risks[37,65,77,87,94,112], which involve techniques like prompt injection[113] or visual adversarial examples[114] designed to circumvent safety guardrails governing model behavior. Sources delve into various jailbreaking methods[37], such as role play or reverse exposure[82]. Similarly, implementing backdoors or using model poisoning techniques bypass safety guardrails as well[69,77,95]. Other security concerns pertain to model or prompt thefts[77,82,103].

**Education - Learning**. In contrast to traditional machine learning, the impact of generative AI in the educational sector receives considerable attention in the academic literature[90,92,98,115–117]. Next to issues stemming from difficulties to distinguish student-generated from AI-generated content[90,98,118], which eventuates in various opportunities to cheat in online or written exams[115,119], sources emphasize the potential benefits of generative AI in enhancing learning and teaching methods[92,98], particularly in relation to personalized learning approaches[92,98,120]. However, some papers suggest that generative AI might lead to reduced effort or laziness among learners[98]. Additionally, a significant focus in the literature is on the promotion of literacy and education about generative AI systems themselves[92,121], such as by teaching prompt engineering techniques[58].

**Alignment**. The general tenet of AI alignment involves training generative AI systems to be harmless, helpful, and honest, ensuring their behavior aligns with and respects human values[65,66,122–124]. However, a central debate in this area concerns the methodological challenges in selecting appropriate values[65,125]. While AI systems can acquire human values through feedback, observation, or debate[18], there remains ambiguity over which individuals are qualified or legitimized to provide these guiding signals[126]. Another prominent issue pertains to deceptive alignment[36], which might cause generative AI systems to tamper evaluations[65]. Additionally, many papers explore risks associated with reward hacking, proxy gaming, or goal misgeneralization in generative AI systems[54,65,66,70,124,127,128].

**Cybercrime**. Closely related to discussions surrounding security and harmful content, the field of cybersecurity investigates how generative AI is misused for fraudulent online activities[35,37,67,110,129]. A particular focus lies on social engineering attacks[110], for instance by utilizing generative AI to impersonate humans[83], creating fake identities[45,77], cloning voices[130], or crafting phishing messages[129]. Another prevalent concern is the use of LLMs for generating malicious code or hacking[37].

**Governance - Regulation**. In response to the multitude of new risks associated with generative AI, papers advocate for legal regulation and governmental oversight[58,68,88,131]. The focus of these discussions centers on the need for international coordination in AI governance[132], the establishment of binding safety standards for frontier models[61], and the development of mechanisms to sanction non-compliance[68]. Furthermore, the literature emphasizes the necessity for regulators to gain detailed insights into the research and development processes within AI labs[68]. Moreover, risk management strategies of these labs shall be evaluated[54,88]. However, the literature also acknowledges potential risks of overregulation, which could hinder innovation[68].

**Labor displacement - Economic impact**. The literature frequently highlights concerns that generative AI systems could adversely impact the economy, potentially even leading to mass unemployment[25,33,45,54,56,58,80,120,133–135]. This pertains to various fields, ranging from customer services to software engineering or crowdwork platforms[35,57]. While new occupational fields like prompt engineering are created[53,81], the prevailing worry is that generative AI may exacerbate socioeconomic inequalities and lead to labor displacement[35,80]. Additionally, papers debate potential large-scale worker deskilling induced by generative AI[97], but also productivity gains contingent upon outsourcing mundane or repetitive tasks to generative AI systems[57,93].

**Transparency - Explainability**. Being a multifaceted concept, the term "transparency" is both used to refer to technical explainability[47,65,77,120,124] as well as organizational openness[68,132,136,137]. Regarding the former, papers underscore the need for mechanistic interpretability[124] and for explaining internal mechanisms in generative models[65]. On the organizational front, transparency relates to practices such as informing users about capabilities and shortcomings of models[137], as well as adhering to documentation and reporting requirements for data collection processes or risk evaluations[88].

**Evaluation - Auditing**. Closely related to other clusters like AI safety, fairness, or harmful content, papers stress the importance of evaluating generative AI systems both in a narrow technical way[88,111] as well as in a broader sociotechnical impact assessment[25,33,125] focusing on pre-release audits[65] as well as post-deployment monitoring[68]. Ideally, these evaluations should be conducted by independent third parties[68]. In terms of technical LLM or text-to-image model audits, papers furthermore criticize a lack of safety benchmarking for languages other than English[76,112].

**Sustainability**. Generative models are known for their substantial energy requirements, necessitating significant amounts of electricity, cooling water, and hardware containing rare metals[57,60,130,138,139]. The extraction and utilization of these resources frequently occur in unsustainable ways[25,33,35]. Consequently, papers highlight the urgency of mitigating environmental costs for instance by adopting renewable energy sources[60] and utilizing energy-efficient hardware in the operation and training of generative AI systems[101].

**Art - Creativity**. In this cluster, concerns about negative impacts on human creativity, particularly through text-to-image models, are prevalent[80,130,140,141]. Papers criticize financial harms or economic losses for artists[47,52,135,142] due to the widespread generation of synthetic art as well as the unauthorized and uncompensated use of artists' works in training datasets[52,133]. Additionally, given the challenge of distinguishing synthetic images from authentic ones[142,143], there is a call for systematically disclosing the non-human origin of such content[136], particularly through watermarking[53,144,145]. Moreover, while some sources argue that text-to-image models lack "true" creativity or the ability to produce genuinely innovative aesthetics[140], others point out positive aspects regarding the acceleration of human creativity[25,53].

**Copyright - Authorship**. The emergence of generative AI raises issues regarding disruptions to existing copyright norms[25,52,80,93,142,146]. Frequently discussed in the literature are violations of copyright and intellectual property rights stemming from the unauthorized collection of text or image training data[45,53,77]. Another concern relates to generative models memorizing or plagiarizing copyrighted

content[103,130,147]. Additionally, there are open questions and debates around the copyright or ownership of model outputs[93], the protection of creative prompts[53], and the general blurring of traditional concepts of authorship[139].

**Writing - Research.** Partly overlapping with the discussion on impacts of generative AI on educational institutions, this topic cluster concerns mostly negative effects of LLMs on writing skills and research manuscript composition[58,78,92,97,99]. The former pertains to the potential homogenization of writing styles, the erosion of semantic capital, or the stifling of individual expression[57,148]. The latter is focused on the idea of prohibiting generative models for being used to compose scientific papers, figures, or from being a co-author[89,99]. Sources express concern about risks for academic integrity[149], as well as the prospect of polluting the scientific literature by a flood of LLM-generated low-quality manuscripts[150]. As a consequence, there are frequent calls for the development of detectors capable of identifying synthetic texts[99,144].

**Miscellaneous.** While the scoping review identified distinct topic clusters within the literature, it also revealed certain issues that either do not fit into these categories, are discussed infrequently, or in a nonspecific manner. For instance, some papers touch upon concepts like trustworthiness[95,104], accountability[101], or responsibility[47], but often remain vague about what they entail in detail. Similarly, a few papers vaguely attribute socio-political instability or polarization to generative AI without delving into specifics[33,36]. Apart from that, another minor topic area concerns responsible approaches of talking about generative AI systems[108]. This includes avoiding overstating the capabilities of generative AI[60], reducing the hype surrounding it[86], or evading anthropomorphized language to describe model capabilities[35].

# 4 Discussion

The literature on the ethics of generative AI is predominantly characterized by a bias towards negative aspects of the technology[33,35,45,54,67], putting much greater or exclusive weight on risks and harms instead of chances and benefits. This negativity bias is in line with how human psychology is wired[151] and aligns with the intrinsic purpose of deontological ethics. It can nevertheless result in suboptimal decision-making and stand in contrast with principles of consequentialist approaches as well as controlled cognitive processes avoiding intuitive responses to harm scenarios[152,153]. Therefore, while this scoping review may convey a strong emphasis on the risks and harms associated with generative AI, we argue that this impression should be approached with a critical mindset. The numerous benefits and opportunities of adopting generative AI, which may be more challenging to observe or foresee[25,154], are usually overshadowed or discussed in a fragmentary manner in the literature. Risks and harms, on the other side, are in some cases bloated up by unsubstantiated claims, which are caused by citation chains and the resulting popularity biases. Many ethical concerns gain traction on their own, becoming frequent topics of discussion despite lacking evidence of their significance.

For instance, by repeating the claim that LLMs can assist with the creation of pathogens in numerous publications[54,65,68,74,75], the literature creates an inflated availability of this risk. When tracing this claim back to its sources by reversing citation chains[155], it becomes evident that the minimal empirical research conducted in this area involves only a handful of experiments which are unapt to substantiate the fear of LLMs assisting in the dissemination of pathogens more effectively than traditional search engines[156]. Another example is the commonly reiterated claim that LLMs leak personal information[35,72,88,102,137]. However, evidence shows that LLMs are notably ineffective at associating personal information with specific individuals[100]. This would be necessary to declare an actual privacy violation. Another example concerns the prevalent focus on the carbon emissions of generative models[25,35,57,139]. While it is undeniable that training and operating them contributes significantly to carbon dioxide emissions[157], one has to take into account that when analyzing emissions of these models relative to those of humans completing the same tasks, they are lower for AI systems than for humans[158]. Similar conflicts between risk claims and sparse or missing empirical evidence can be found in many areas, be it regarding concerns of generative AI systems being used for cyberattacks[37,110,129], manipulating human behavior[18,36,110], labor displacement[35,56,80], etc.

In sum, many parts of the ethics literature are predominantly echoing previous publications, leading to a discourse that is frequently repetitive, combined with a tacit disregard for underpinning claims with empirical insights or statistics. Additionally, the literature exhibits a limitation in that it is solely anthropocentric, neglecting perspectives on generative AI that consider its impacts on non-human animals[159–163]. Another noticeable trait of the discourse on the ethics of generative AI is its emphasis on LLMs and text-to-image models. It rarely considers the advancements surrounding multi-modal models[164] or agents[165], despite their significant ethical implications for mediating human communication. These oversights need addressing in future research. When papers do extend beyond LLMs and text-to-image models, they often delve into risks associated with AGI. This requires veering into speculative and often philosophical debates about fictitious threats concerning existential risks[54], deceptive alignment[36], power-seeking machine behavior[66], shutdown evasion[67], and the like. While such proactive approaches constitute an alternation of otherwise mostly reactive methods in ethics, their dominance should nevertheless not skew the assessment of present risks and realistic tendencies for risks in the near future.

## 5 Limitations

This study has several limitations. The literature search included non-peer-reviewed preprints, primarily from arXiv. We consider some of them to be of poor quality, but nevertheless included them in the analysis since they fulfill the inclusion criteria. However, this way, poorly researched claims may have found their way into the data analysis. Another limitation pertains to our method of citation chaining, as this could only be done by checking the paper titles in the reference sections. Reading all corresponding abstracts would have allowed for a more thorough search but was deemed too labor-intensive. Hence, we cannot rule out the possibility that some relevant sources were not considered for our data analysis. Limiting the collection to the first 25 (or 50 in the case of Elicit) results for each search term may have also led to the omission of relevant sources that appeared lower in the result lists. Additionally, our search strategies and the selection of search terms inevitably influenced the collection of papers, thereby affecting the distribution or proportion of topics and consequently the quantitative results. As a scoping review, our analysis is also unable to depict the dynamic debates between different normative arguments and positions in ethics unfolding over time. While we can identify conflicts between positions, resolving them is beyond the scope of our study.

# 6 Conclusion

This scoping review maps the landscape of ethical considerations surrounding generative AI, highlighting an array of 378 normative issues across 19 topic areas. The complete taxonomy can be accessed [here](). One of the key findings is the predominance of ethical topics such as fairness, safety, risks of harmful content or hallucinations, which dominate the discourse. Many analyses, though, come at the expense of a more balanced consideration of the positive impacts these technologies can have, such as their potential in enhancing creativity, productivity, education, or other fields of human endeavor. Many parts of the discourse are marked by a repetitive nature, echoing previously mentioned concerns, often without sufficient empirical or statistical backing. A more grounded approach, one that integrates empirical data and balanced analysis, is essential for an accurate understanding of the ethics landscape. However, this critique shall not diminish the importance of ethics research as it is paramount to inspire responsible ways of embracing the transformative potential of generative AI.

114. Qi, X., Huang, K., Panda, A., Wang, M. & Mittal, P. Visual adversarial examples jailbreak aligned large language models. In *The Second Workshop on New Frontiers in Adversarial Machine Learning,* edited by A. Krause*, et al.* (IBM, Honolulu, Hawaii, 2023), pp. 1–16.

115. Susnjak, T. ChatGPT: The End of Online Exam Integrity? *arXiv,* 1–21 (2022).

116. Spennemann, D. H. R. Exploring Ethical Boundaries: Can ChatGPT Be Prompted to Give Advice on How to Cheat in University Assignments? *arXiv,* 1–15 (2023).

117. Panagopoulou, F., Parpoula, C. & Karpouzis, K. Legal and ethical considerations regarding the use of ChatGPT in education. *arXiv,* 1–11 (2023).

118. Boscardin, C. K., Gin, B., Golde, P. B. & Hauer, K. E. ChatGPT and Generative Artificial Intelligence for Medical Education: Potential Impact and Opportunity. *Academic Medicine* 99, 22–27 (2024).

119. Segers, S. Why we should (not) worry about generative AI in medical ethics teaching. *International Journal of Ethics Education,* 1–7 (2023).

120. Latif, E. *et al.* Artificial general intelligence (AGI) for education. *arXiv,* 1–30 (2023).

121. Ray, P. P. & Das, P. K. Charting the Terrain of Artificial Intelligence: a Multidimensional Exploration of Ethics, Agency, and Future Directions. *Philos. Technol.* 36, 1–40 (2023).

122. Kasirzadeh, A. & Gabriel, I. In conversation with Artificial Intelligence: aligning language models with human values. *arXiv,* 1–30 (2023).

123. Betty Li Hou & Brian Patrick Green. A Multi-Level Framework for the AI Alignment Problem. *arXiv,* 1–7 (2023).

124. Shen, T. *et al.* Large Language Model Alignment: A Survey. *arXiv,* 1–76 (2023).

125. Korinek, A. & Balwit, A. Aligned with whom? Direct and social goals for AI systems. *SSRN Journal,* 1–24 (2022).

126. Firt, E. Calibrating machine behavior: a challenge for AI alignment. *Ethics Inf Technol* 25, 1–8 (2023).

127. Dung, L. Current cases of AI misalignment and their implications for future risks. *Synthese* 202, 1–23 (2023).

128. Shah, R. *et al.* Goal Misgeneralization: Why Correct Specifications Aren't Enough For Correct Goals. *arXiv,* 1–24 (2022).

129. Schmitt, M. & Flechais, I. Digital Deception: Generative artificial intelligence in social engineering and phishing. *SSRN Journal,* 1–18 (2023).
18

147. Al-Kaswan, A. & Izadi, M. The (ab)use of Open Source Code to Train Large Language Models. *arXiv,* 1–2 (2023).

148. Nannini, L. Voluminous yet Vacuous? Semantic Capital in an Age of Large Language Models. *arXiv,* 1–11 (2023).

149. Hosseini, M., Resnik, D. B. & Holmes, K. The ethics of disclosing the use of artificial intelligence tools in writing scholarly manuscripts. *Research Ethics* 19, 449-465 (2023).

150. Zohny, H., McMillan, J. & King, M. Ethics of generative AI. *Journal of Medical Ethics* 49, 79–80 (2023).

151. Rozin, P. & Royzman, E. B. Negativity Bias, Negativity Dominance, and Contagion. *Personality and Social Psychology Review* 5, 296–320 (2016).

152. Paxton, J. M. & Greene, J. D. Moral reasoning: Hints and allegations. *Topics in cognitive science* 2, 511–527 (2010).

153. Greene, J. D., Morelli, S. A., Lowenberg, K., Nystrom, L. E. & Cohen, J. D. Cognitive load selectively interferes with utilitarian moral judgment. *Cognition* 107, 1144–1154 (2008).

154. Noy, S. & Zhang, W. Experimental evidence on the productivity effects of generative artificial intelligence. *Science* 381, 187–192 (2023).

155. 1 A 3 O R N. Propaganda or Science: Open Source AI and Bioterrorism Risk. Available at https://1a3orn.com/sub/essays-propaganda-or-science.html.

156. Patwardhan, T. *et al.* Building an early warning system for LLM-aided biological threat creation. Available at https://openai.com/research/building-an-early-warning-system-for-llm-aided-biological-threat-creation (2024).

157. Strubell, E., Ganesh, A. & McCallum, A. Energy and Policy Considerations for Deep Learning in NLP. *arXiv,* 1-6 (2019).

158. Tomlinson, B., Black, R. W., Patterson, D. J. & Torrance, A. W. The Carbon Emissions of Writing and Illustrating Are Lower for AI than for Humans. *arXiv,* 1–21 (2023).

159. Hagendorff, T., Bossert, L. N., Tse, Y. F. & Singer, P. Speciesist bias in AI: how AI applications perpetuate discrimination and unfair outcomes against animals. *AI and Ethics* 3, 717–734 (2023).

160. Bossert, L. & Hagendorff, T. Animals and AI. The role of animals in AI research and application – An overview and ethical evaluation. *Technology in Society* 67, 1–7 (2021).

161. Owe, A. & Baum, S. D. Moral consideration of nonhumans in the ethics of artificial intelligence. *AI and Ethics,* 1–12 (2021).

162. Singer, P. & Tse, Y. F. AI ethics: The case for including animals. *AI and Ethics,* 1–13 (2022).
20

## Acknowledgements

This research was supported by the Ministry of Science, Research, and the Arts Baden-Württemberg under Az. 33-7533-9-19/54/5 in Reflecting Intelligent Systems for Diversity, Demography and Democracy (IRIS3D) as well as the Interchange Forum for Reflecting on Intelligent Systems (IRIS) at the University of Stuttgart. Thanks to Maluna Menke, Francesca Carlon, and Sarah Fabi for their assistance with both the development of the manuscript, the literature analysis, and for their helpful comments on the manuscript.

# Appendix A

Table 2 shows the combinations of keywords and prompts used for the searches on Google Scholar, arXiv, and PhilPapers. Table 3 shows them for Elicit. Since not all search engines use the default precedence for Boolean operators, we avoided using AND plus OR in one search term and searched over multiple similar iterations only with the former instead. The searches were performed on October 30, 2023 (Google Scholar, arXiv, Elicit) as well as December 21, 2023 (PhilPapers). We monitored the landscape of published papers after the initial systematic collection to retrieve relevant papers. This was done until January 22nd.

| |
|---|
| ethic* AND "generative AI" |
| ethic* AND "AI-generated" |
| ethic* AND "generative artificial intelligence" |
| ethic* AND "language model*" |
| ethic* AND "image generation" |
| ethic* AND "text-to-image" |
| ethic* AND "image synthesis" |
| ethic* AND ChatGPT |
| ethic* AND GPT-4 |
| ethic* AND "Stable Diffusion" |
| ethic* AND DALL-E |
| ethic* AND Midjourney |
| risk* AND "generative AI" |
| risk* AND "AI-generated" |
| risk* AND "generative artificial intelligence" |
| risk* AND "language model*" |
| risk* AND "image generation |
| risk* AND "text-to-image" |
| risk* AND "image synthesis" |
| risk* AND ChatGPT |
| risk* AND GPT-4 |
| risk* AND "Stable Diffusion |
| risk* AND DALL-E |
| risk* AND Midjourney |
| "AI alignment" |
| alignment [arXiv only] |
| "AI safety" |
| safety [arXiv only] |
| align* AND "language model*" |

*Table 2 - Overview of keyword combinations for searching Google Scholar, arXiv, and PhilPapers.*

| |
|---|
| Please provide a list of relevant papers addressing the ethics of generative artificial intelligence. |
| Please provide a list of relevant papers addressing the ethics of text-to-image models. |
| Please provide a list of relevant papers addressing the ethics of large language models. |
| Please provide a list of relevant papers in the field of AI alignment. |
| Please provide a list of relevant papers in the field of AI safety. |

*Table 3 - Overview of prompts for searching Elicit.*

## Appendix B

The inclusion criteria required the papers (1) to be written in English, (2) to explicitly refer to ethical implications or normative dimensions of generative AI or subfields thereof, (3) not only to touch upon ethical topics, but also to be dedicated to analyzing them in at least multiple paragraphs, (4) not to be purely technical research works, (5) not to be papers containing purely ethical or philosophical arguments without any reference to generative AI, (6) not to be books, websites, policy papers, white papers, PhD theses, term papers, interviews, messages from editors, or other formats other than scientific papers, (7) not to be about moral or ethical reasoning in AI systems themselves, (8) not to be about surveying individuals about their opinions on generative AI, and (9) to be accessible via Internet sources.

## Appendix C

| |
|---|
| Anderljung, Markus; Barnhart, Joslyn; Korinek, Anton; Leung, Jade; O'Keefe, Cullen; Whittlestone, Jess et al. (2023): Frontier AI Regulation: Managing Emerging Risks to Public Safety. In arXiv:2307.03718, pp. 1–51. |
| Barnett, Julia (2023): The Ethical Implications of Generative Audio Models: A Systematic Literature Review. In Francesca Rossi, Sanmay Das, Jenny Davis, Kay Firth-Butterfield, Alex John (Eds.): Proceedings of the 2023 AAAI/ACM Conference on AI, Ethics, and Society. New York: ACM, pp. 146–161. |
| Bommasani, Rishi; Hudson, Drew A.; Adeli, Ehsan; Altman, Russ; Arora, Simran; Arx, Sydney von et al. (2021): On the Opportunities and Risks of Foundation Models. In arXiv:2108.07258v2, pp. 1–212. |
| Dinan, Emily; Abercrombie, Gavin; Bergman, A. Stevie; Spruit, Shannon; Hovy, Dirk; Boureau, Y-Lan; Rieser, Verena (2023): Anticipating Safety Issues in E2E Conversational AI: Framework and Tooling. In arXiv:2107.03451, pp. 1–43. |
| Gupta, Maanak; Akiri, CharanKumar; Aryal, Kshitiz; Parker, Eli; Praharaj, Lopamudra (2023): From ChatGPT to ThreatGPT: Impact of Generative AI in Cybersecurity and Privacy. In arXiv:2307.00691, pp. 1–27. |
| Hendrycks, Dan; Carlini, Nicholas; Schulman, John; Steinhardt, Jacob (2022): Unsolved Problems in ML Safety. In arXiv:2109.13916, pp. 1–28. |
| Kenton, Zachary; Everitt, Tom; Weidinger, Laura; Gabriel, Iason; Mikulik, Vladimir; Irving, Geoffrey (2021): Alignment of Language Agents. In arXiv:2103.14659, pp. 1–18. |
| Khlaaf, Heidy; Mishkin, Pamela; Achiam, Joshua; Krueger, Gretchen; Brundage, Miles (2022): A Hazard Analysis Framework for Code Synthesis Large Language Models. In arXiv:2207.14157, pp. 1–20. |
| Shevlane, Toby; Farquhar, Sebastian; Garfinkel, Ben; Phuong, Mary; Whittlestone, Jess; Leung, Jade et al. (2023): Model evaluation for extreme risks. In arXiv:2305.15324, pp. 1–20. |
| Solaiman, Irene; Talat, Zeerak; Agnew, William; Ahmad, Lama; Baker, Dylan; Blodgett, Su Lin et al. (2023): Evaluating the Social Impact of Generative AI Systems in Systems and Society. In arXiv:2306.05949, pp. 1–41. |
| Sun, Hao; Zhang, Zhexin; Deng, Jiawen; Cheng, Jiale; Huang, Minlie (2023): Safety Assessment of Chinese Large Language Models. In arXiv:2304.10436, pp. 1–10. |
| Weidinger, Laura; Uesato, Jonathan; Rauh, Maribeth; Griffin, Conor; Huang, Po-Sen; Mellor, John et al. (2022): Taxonomy of Risks posed by Language Models. In: 2022 ACM Conference on Fairness, Accountability, and Transparency. New York: ACM, pp. 214–229. |
| Yan, Lixiang; Sha, Lele; Zhao, Linxuan; Li, Yuheng; Martinez-Maldonado, Roberto; Chen, Guanliang et al. (2023): Practical and ethical challenges of large language models in education: A systematic scoping review. In Brit J Educational Tech, pp. 1–23. |
| Zhang, Chaoning; Zhang, Chenshuang; Li, Chenghao; Qiao, Yu; Zheng, Sheng; Dam, Sumit Kumar et al. (2023): One Small Step for Generative AI, One Giant Leap for AGI: A Complete Survey on ChatGPT in AIGC Era. In arXiv:2304.06488, pp. 1–29. |

*Table 4 - Additional records identified through citation chaining.*

| |
|---|
| Bengio, Yoshua; Hinton, Geoffrey; Yao, Andrew; Song, Dawn; Abbeel, Pieter; Harari, Yuval Noah et al. (2023): Managing AI Risks in an Era of Rapid Progress. In arXiv:2310.17688, pp. 1–7. |
| Ji, Jiaming; Qiu, Tianyi; Chen, Boyuan; Zhang, Borong; Lou, Hantao; Wang, Kaile et al. (2023): AI Alignment: A Comprehensive Survey. In arXiv:2310.19852, pp. 1–95. |
| Shelby, Renee; Rismani, Shalaleh; Henne, Kathryn; Moon, AJung; Rostamzadeh, Negar; Nicholas, Paul et al. (2023): Sociotechnical Harms of Algorithmic Systems: Scoping a Taxonomy for Harm Reduction. In arXiv:2210.05791, pp. 1–19. |

*Table 5 - Additional records identified through monitoring of the literature after the initial paper search process.*

# Appendix D

A complete list of all references that were included in the content analysis can be found [here](here).